# High-throughput nanoparticle analysis in a FEG-SEM using an inexpensive multi-sample STEM-ADF system

M.J.Lagos[1,2], P.C.da Silva[1], D.Ugarte[2] [*]

[1]Laboratório Nacional de Luz Síncrotron, C.P. 6192,
13084-971 Campinas SP, Brazil.
[2]Instituto de Física "Gleb Wataghin", Universidade Estadual de Campinas,
UNICAMP,
13083-970 Campinas SP, Brazil.

*To whom correspondence should be addressed. E-mail:dmugarte@ifi.unicamp.br

**Abstract**

Nanotechnology research requires the routine use of characterization methods with high spatial resolution. These experiments are rather costly, not only from the point of view of the expensive microscopes, but also considering the need of a rather specialized equipment operator. Here, we describe the construction of an inexpensive and simple device that allows the analysis of nanoparticle in a FEG-SEM; images can be generated at high magnifications (ex. x500.000) and with nanometric resolution. It is based on the acquisition of transmitted electrons annular dark field (TE-ADF) signal; the systems can carry up to 16 TEM samples and, it is compatible with SEM sample exchange air-lock. Performance test have shown the measured ADF signal showed the atomic number and thickness dependence for transition metal nanoparticle about 10 nm in diameter. Also, the signal quality is high enough that the determination of the histogram of size distribution can be performed using a conventional image processing software, for gold particles in the range of 2-10 nm in diameter. The developed ADF device allows a much faster and cheaper high spatial resolution imaging of nanoparticle samples for routine morphological characterization and, provides an invaluable high throughput tool for an efficient sample screening.

# 1. Introduction

Nanoscience and nanotechnology research requires routine use of characterization methods with high spatial resolution to analyze samples at the nanometric scale both for morphological and spectroscopic studies. For example, during the improvement of a synthesis protocol a reasonable large number of nanoparticle or nanowire samples must be imaged to gather different parameters (size, morphology, etc.) not only for the average value but to get the full distribution function. This actually represents a rather expensive fraction of nanosystem research, not only from the point of view of the high cost of instruments purchase and operation, but also considering the need of experienced and rather specialized equipment operator.

Among the most widely used experimental approaches, Transmission Electron Microscopy (TEMs) represents a clear example of the ideas discussed above. TEM and Scanning TEM (STEM) methods show excellent spatial resolution (angstroms), but their use requires a highly trained operator and the studies may be rather time consuming. Interpretation of TEM images adds additional complications, because, in fact, it may be rather tricky. We must keep in mind that image contrast is originated by dynamical diffraction effects that generate an unintuitive dependence of intensities with electron beam properties (size, energy width, angular opening, etc.) and the sample atomic arrangement, crystal orientation, thickness, etc. Recently, electron microscopes have made a huge progress and sub-angstrom resolution is at present "off the shelf"; this opens new opportunities for cutting edge experiments [1]. Although, advanced control and user interface seem to render these instruments much friendly, the need of qualification and experience operators still raises a reasonable barrier for the regular use of even the conventional TEM in many research groups.

Scanning Electron Microscopes (SEMs) represent the most popular and wide-spread used electron-beam based instrument; Field-Emission-Gun (FEG) equipped SEMS can be able to attain nanometric resolution using the secondary electron (SE) signal. SEMs are easy to use in an intuitive way, images are much easier to interpret and, in addition, they may have a high throughput (quicker alignment of the electron optics, image acquisition and sample exchange). The simpler use and high versatility of SEM microscopes are their stronger attribute. SE images have been intensively used for carbon nanotubes and nanowires [2], but are not necessary well suited to analyze nanoparticles with diameter < 20 nm. Resolution loss associated with size reduction (edge or border effects [2]) and small signal difficult NP imaging; lowering the beam energy may reduce these effects, however this increases Boersh effect (repulsion between electron inside the narrow beam [2,3]) broadening electron beam size.

In recent years, the popularization of Scanning Transmission Electron Microscopes (STEMs) has rendered Annular Dark Field (ADF) images a standard tool to analyze materials with nanometric or even atomic resolution [4]. These images show high contrast and are easier to interpret than the diffraction-based-ones obtained by TEM. In fact, intensity can be considered in first approximation proportional to mean atomic number (Z) and thickness of the analyzed region. Then, as a natural evolution, modern FEG-SEMs have been adapted to analyze conventional TEM thin samples by installing transmitted electron (TE) detectors for both Brigth Field (BF) and ADF imaging techniques [5-8]. These new detectors exploit the narrow FEG-SEM electron probes, but the inherent lower electron energy (25-30 kV) somewhat limits the technique application to very thin- or low atomic number samples (ex. biological).

In this work, we describe the fabrication of a conceptually simple and inexpensive TE-ADF device with multi-sample capability operating inside a SEM, where the TE signal is measured using the conventional secondary electron detector. The fabricated holder allows an easy and quick generation of high quality ADF STEM images of nanoparticles with nanometric resolution. This system is compatible with SEM sample exchange air-lock and, high throughput is guaranteed by the fact that it can carry up to 16 TEM samples.

## 2. STEM-ADF sample holder: design and construction

In a scanning transmission electron microscope, a fine electron probe is formed by using a system of strong electron lens to demagnify a small electron source and, is scanned over a transmission sample in a two dimensional raster (Fig. 1). The transmitted electrons (TE) are scattered at different angles. If the detector collects the primarily transmitted beam (low-angle scattered electrons) a bright field (BF) image is formed. But, detection of all transmitted electrons scattered around the central beam gives an ADF image [3,4]. In particular, ADF STEM imaging mode provides better image resolution due to the collection of high-angle elastic scattered electrons.

We have used ADF concept to build a device operated in a scanning electron microscope (SEM). A scheme of our simple STEM-ADF device and its operating mode are shown in Figure 2a. Basically, the fabricated device is cylindrical in shape (20 mm in length and 12 mm in diameter Fig. 2b). It holds the conventional 3 mm TEM grids; a drilled cap is used to fix TEM grids and avoid sample drift. A hole below the TEM sample (1 mm in diam.) allows the transmission of forward scattered electrons (TE signal). The TEs impinge on an electron-conversion layer, yielding low-energy electrons (SE) which are collected by the conventional Everhart-Thornley (ET) SE detector [6-8]. In this way, the operation of the STEM device does not require any modification of the original SEM; the STEM operation remains as simple and user-friendly as the SEM itself. The STEM device is mounted on the conventional stage specimen holder and introduced through the microscope air-lock. Once inside the microscope, the holder is elevated with the Z-axis control until the sample is at a working distance of about 5 mm from the final lens (this is the shorter working distance we estimated in our system in order to avoid possible damage to microscope parts). In order to extract the ADF contribution from the TE beam, a hole is drilled on the conversion layer below the sample position. In this way, the low angle scattered electrons (STEM-BF signal) are absorbed inside the hole (Fig. 2a), and the resulting SE signal is only originated from high angle scattered electrons striking on the conversion layer, what represents effectively an ADF signal.

Concerning the materials used to fabricate the STEM device, it was built in graphite in order to reduce the SE emission by backscattered electrons around the sample and, also, to absorb efficiently the low angle scattered TE below it. Although the conceptual design is simple and straightforward, the electron conversion surface must be optimized to transform the TE-ADF electrons into SE with high efficiency in order to get a reasonable signal noise ratio from the tiny nanoparticle samples. It is well known that SE emission increases with increasing both specimen tilt angle and atomic number [2]. To fulfill the requirement of grazing incidence electron trajectories and, at the same time, allowing the escape of SE in direction of the Everhart-Thornley detector, we have built a concave hemi-cylindrical surface whose axis is perpendicular to the electron beam (see Fig. 2a). The TE beam travels very close to the cylindrical surface and, in this way, a significant part of high angle scattered transmitted electrons strike on the conversion surface at very small angles. The STEM device must be rotated along the vertical axis in order to direct the surface opening to the microscope SE detector. Concerning the electron converting layer, it was prepared by evaporating a 150 nm gold (high atomic number element) film over the graphite cylindrical surface. The pure metal surface yielded an ADF signal with good signal to noise ratio (S/N) images. The ADF signal was further improved by gluing a magnesium oxide layer over the gold surface; following old recipes on preparation of surfaces with high SE yield [2,9]. This improved the S/N of approximately a factor 3.

In order to allow a higher throughput we further completed our design for a multi-sample system. Due to the simple geometry of the mechanical design, it could be easily extended to generate a larger cylindrical carrousel that keeps all the geometrical characteristics of the electron beam TE-SE conversion (the cylindrical conversion surface became a toroidal surface). This larger STEM device is about 50 mm in diameter and 25 mm in height; it compatible with the normal SEM sample exchange air-lock and can receive up to 16 TEM grids (Fig. 2c). To reduce microscope pumping time, the device body was fabricated in Al, because a large mass of graphite takes long time to degas and sample exchange becomes slower.

## 3. Performance Tests

We have tested our ADF device in a SEM (JSM-6330F) equipped with a cold field emission gun. This SEM uses a conventional Everhart-Thornley detector and is not equipped with in-lens electron detection. No modification was realized on the microscope and, all ADF images presented in this work have used the microscope electronics and control as processing a normal SE image while using the STEM-ADF holder.

In order to make an evaluation of the STEM-ADF performance, we must first compare conventional SE images and the ADF ones derived from the same microscope. However, as the physical processes originating these signals are very different, we must optimize the imaging conditions for each mode independently. In these terms, the accelerating voltage used during image acquisition was 5 for both SE in order to avoid loss of resolution due to the edge effect [2] and, 30 KV for ADF mode in order to get the highest possible penetration power and minimize beam broadening. Other important microscope parameters such as working distance (5 mm), objective lens aperture (30 μm), probe current (~$1 \times 10^{-10}$ A), acquisition time (~20 s), magnification (x500000), pixel size (0.18 nm), sampling (1280x1024 pixels) and dynamical range (8 bits / 256 gray level tones) were identical for both imaging modes.

Our initial test sample was a TEM grid containing 20-40 nm gold NPs supported on a carbon film. This sample is a well behaved and generates images that are easy to understand/modelize. Fig. 3 shows images of a small cluster of gold NPs imagined by both modes. A simple visual comparative analysis reveals clearly that ADF image presents higher contrast and sharper edges but with a slight noise increase, when compared to SE image. In order to make a quantitative analysis of the resolution comparison, we have calculated the Fourier Transforms (FT) of the corresponding ADF and SE images (insets in Fig.3) [2]. A visual analysis of both Fourier transforms indicates that ADF FT displays information for much larger momentum vectors, what can be understood as a substantial improvement of the resolution in ADF mode. The maximum spatial frequency measured from FTs spectra for these samples were about 5 and 8 nm in ADF and SE image, respectively. It is possible that that some carbon contamination deposition has contributed to slightly reduce resolution; in fact sample contamination may become critical when observing samples at high magnification (x300.000-500.000).

A criterion that is frequently used to estimate the image resolution at an interface is the concept of shadow (edge resolution), which defines resolution as the distance between points corresponding to 25% and 75% of the total intensity variation [3]. To perform this measurement, we have used gold nanorods as test sample in order to get a regular step intensity profile at the longer sides of a gold nanorod. Fig.3c shows an ADF image and, an integrated intensity obtained across the nanorod edge. The analysis of this profile yields an edge resolution of (0.90 ± 0.03) nm, what represents a reasonable improvement when compared to microscope SE nominal resolution (1.5 nm).

## 4. Discussion and Further Applications

The fabricated STEM-ADF system allows an easy and quick generation of high quality ADF images of NPs at highest magnifications (X500.000), rendering its operation much more efficient for NP characterization than when using the conventional SE mode. As further test of application we have analyzed nickel-, iron-, silver-, lead thelurade- and gold NPs, as well as, carbon nanotubes.

It is interesting to note that ADF imaging may also provide additional structural information of the NPs. In Fig. 3b, the ADF intensity is almost equivalent all over the triangle-shaped NP, indicating that this NP is probably of uniform thickness. The spherical NPs display a ring-like contrast (intensity increase of at the edges). Considering that this TE-ADF signal was obtained with rather low electron energy (30 KeV), the intensity reduction at the particle center is due to electron absorption at the thicker NP regions (the mean free path for electron total scattering is ~ 3 nm in gold). This example clearly indicates that SEM TE-ADF imaging provides much more information, allowing the discrimination between NPs with different morphologies (flat triangle from pyramid and almost spheres from discs). However, we must also emphasize that these images clearly reveal the limitations of acquiring TE-ADF signals using SEM beams and, that even ~20 nm Au NPs must be considered a rather thick sample for this imaging mode.

To test whether the STEM-ADF holder can generate images of more challenging NP samples, we have analyzed smaller particles that are made of a lower atomic number material (transition metal, Ni and Fe). Fig. 4a shows clearly that high contrast ADF images are generated when observing 8-12 nm nickel NPs deposited on thin carbon film (4-5 nm in thickness). The NP contrast is more disc-like, in contrast with the ring-like one observed for the 30 nm Au ones; this indicates that the SEM electron beam is not strongly absorbed by these particles. This conclusion is confirmed by the analysis of intensity profiles across a line of NPs (Fig. 4b), each particle is represented by a broad peak with the maximum intensity at the NP center (thickness maximum). A more detailed analysis reveals clearly that NPs with the same size (~ 12 nm) generate identical intensity variation (Fig. 4b, curve 1), while NPs with different size (~ 12 and 10 nm) presents a significant variation of signal contrast (Fig. 4b, curve 2, 3). These intensity variations are statistical meaningful (5 times larger than statistical noise), confirming that, using the developed STEM-ADF device, a detailed intensity analysis can provide information on size variations for the case of transition metal NP in the 10 nm size range.

One of the most important features of STEM-ADF signal is the strong dependence on sample atomic number. In this way, ADF imaging mode has become an invaluable tool to get high contrast images of high atomic number particles deposed/immersed on/in light-atom film/matrix (ex. a catalyst made of Pt particle on Zeolite or carbon) [10]. Then, it is necessary to evaluate the performance of our TE-ADF device concerning the signal generated by NP of different chemical composition. To experimentally test this issue, we have acquired ADF images of a TEM sample containing silver and iron NPs of 12-15 in diameter deposited on a lacey carbon film (see Fig.4c). A quick visual analysis of the ADF image contrast reveals that two kinds of NPs are present (or two different levels of signal intensity). A more quantitative analysis can be performed comparing intensity profiles including neighboring NPs with similar size and different contrast. Fig. 4d (curves 1-3) display the profiles of selected image regions; as expected each profile displays two broad peaks. It can be clearly noted that the intensity difference between the particles is significant (30-40 % difference in the intensity maximum). If we assume that ADF signal is proportional to $Z^{2/3}$ [4], then it is possible to roughly estimate, signal ratio for NPs of similar size and different chemical compositions. Thus, it is expected that silver (Z = 47) and iron (Z = 26) NPs will generate an ADF signal ratio of ~1.5; what agrees very well with the values derived from the analyses of several ADF images (see Fig. 4d). Summarizing, the STEM-ADF device allows in fact a reasonable identification of NP with different atomic numbers ($Z_1$, $Z_2$) in favorable cases (for our test sample $\Delta Z$ ~20 and $Z_1/Z_2$ ~1.8). However we must keep in mind that but this simple and quick analysis should be used with caution, as beam absorption effects are very important for the acceleration voltages currently used in SEM.

As natural matter of facts, we must make more stringent practical questions concerning on TE-ADF device performance: a) which is the smallest NP size that can be measured in the ADF image? ; b) is it possible to generate valuable data to derive easily the size distribution of a NP sample? To analyze this point, we observed a sample containing gold NPs dispersed on a carbon film with typical diameter below 10 nm and a rather broad size distribution (i.e. simulating an unsuccessful synthesis experiment, see Fig. 5a). This image was computer processed in conventional way (Digital Micrograph software) to automatically identify particles and yield the diameter distribution (Fig 5b). The diameters histogram displays a large spread, indicating the existence of Au NPs with diameters between 2 and 8 nm and a mean diameter of ~ 3.5 nm. This demonstrates that the developed STEM-ADF image coupled to a routine image processing indeed provide a quick tool to evaluate the NP sample quality; for example if narrow distribution is a target, the sample imaged in Fig. 5a can be disconsidered for any further TEM work. Nevertheless, we must keep in mind that STEM-ADF images show high contrast but are much noisier than conventional TEM images. Then, a subsequent rigorous TEM measurement of the best samples should be realized. In this way, our STEM-ADF imaging device does not aim substitute to TEM studies, but it shows the great capability of making quick screening studies of NPs samples [11].

Finally, we will discuss some more general applications of the STEM-ADF imaging in nanostructured materials. Fig. 5c displays an image of a coating composed by the stacking of layers formed by layers of silicon oxide and, a nanostructured film generated by the deposition of TePb nanoparticles (6-7 nm in diameter, generated by laser-evaporation). The image contrast follows the strong variation of chemical composition between layers, the individual TePb NPs can be clearly observed at the thinnest regions of the sample (upper left corner). The interlayer spacing and some deviation of planarity can be identified and quantified from the image. As regards light element systems, Fig. 5d shows the image of multiwalled carbon nanotube of about 15 nm in diameter. As the atomic number of carbon ($Z= 6$) is rather low, the ADF signal is significantly reduced; however, the empty inner space of the nanotube (that actually represents a thickness reduction) can be easily detected, as revealed by the intensity profile shown in inset of Fig. 5d. The ADF carbon nanotube image displays a rather fuzzy and corrugated nanotube border that contrasts with the sharp and flat interface generated by the curved graphite layers in TEM images. This phenomenon reveals the rather low signal/noise ratio generated by the simple STEM-ADF device at the thinnest region of a low atomic number sample. This suggests that nanotube diameter measurements may be somewhat underestimated when using the STEM –ADF device.

## 5. Summary

We have designed and built an inexpensive and conceptually simple device that allows the acquisition of ADF transmitted electrons images in a FEG-SEM**,** using conventional TEM thin samples. The device converts the STEM-ADF beam in secondary electrons using a surface conversion layer; the signal is detected using the SEM conventional Everhart-Thornley detector. Then, this device does not require any modification of the microscope and, ADF images are acquired using the microscope electronics and user interface. The holder is compatible with SEM sample exchange air-locks and, high throughput sample analysis is guaranteed by the fact that it can carry up to 16 TEM samples.

We have tested our ADF device in a SEM equipped with a cold field emission gun. The generated images have shown high contrast and, in particular, subnanometric resolution (0.9 nm) was achieved when imaging gold nanorods (30 nm in diameter). Although the simplicity of the developed ADF sample holder, we have shown above that the produced dark field signal display the expected atomic number and thickness dependence for transition metal NP particles samples in the 10 nm range. The ADF images display the necessary quality to be used in the automatic calculation of histogram of size distribution; this fact was tested measuring a sample containing Au NP in the range of 2-10 nm in diameter and, using the conventional image processing software. Briefly, the TE-ADF device allow a much faster and cheaper routine morphological analysis of particles deposed on TEM grids; this performance allows a very efficient screening/selection of the best NP samples for further studies.

As mentioned earlier, it is very frequent that nanomaterial research programs require the intense use of TEM to routinely analyze many NPs samples (screening). Our electron microscopy user facility has been challenged by this kind of TEM high demand for routine characterization (mean size, sample uniformity, average shape, etc.), which is very time consuming due to inherent difficulties of TEM methods. In fact, we have met serious difficulty to provide a minimal flux of microscope access for the users guaranteeing an appropriate dynamics for user research. The operation of STEM-ADF device described here has provided a quick morphological analysis of nanoparticle with high spatial resolution for the everyday facility operation and allowed the optimization of the microscope use.

## Acknowledgments

The authors would like to thank the financial support of CNPq, FAPESP and LNLS. We are also thankful to D. Zanchet for providing NP samples and L.C. Barbosa for the multilayer sample.

Figure Captions

Figure 1. Scheme of the transmitted electron signal in Scanning Transmission Electron Microscopy. Two main imaging modes can be implemented: bright field (BF) collecting low angle scattered electrons and annular dark field (ADF) collecting the high angle scattered ones.

Figure 2. (a) Scheme of the designed and operating of the TE-ADF system (see text for explanations. (b) Single sample ADF-STEM device (height 20 mm); (c) carrousel-type cylindrical multi-sample TE-ADF device that accepts 16 TEM grids simultaneously.

Figure 3. Micrographs of gold nanoparticles (NPs) obtained by SE mode (a) and, ADF mode (b). Insets: corresponding image Fourier Transforms (FT). Note that the ADF FT shows information for longer momentum vectors, indicating that higher resolution is obtained. (c) ADF image of gold nanorods and, an intensity profile across the particle edge (d). The profile intensity was obtained by averaging over 100 intensity lines acquired inside the area indicated by a rectangle in (c).

Figure 4. (a) ADF image of 8-12 nm nickel NPs supported on a thin carbon film. (b) Integrated line profiles obtained for three consecutives NPs; the numbered white lines in (a) identify the profile positions in the image. (c) ADF image of sample containing NP of different materials silver and iron (12-15 nm in diameter); d) intensity profile extracted from (c); the numbered arrows identify the NP positions where intensity profiles were obtained. Each profile was obtained by summing over 10 intensity line; the curves have been vertically shifted to render visualization easier (see text for explanations).

Figure 5. (a) ADF image of gold NPs suspended on thin carbon film. (b) Size distribution histogram of NPs shown in (a). Note that histogram reveals a wide NP size distribution (2-8 nm). (c) ADF image of sample composed of multilayers of 6-7 nm TePb NPs and SiO. Gray and white bands in image correspond to SiO and TePb NPs layers, respectively. NPs are easily observed at the thinner regions of the sample (left side of the image). (d) ADF image of carbon nanotube of about 15 nm in diameter. Inset: intensity profile derived from the rectangular region indicated in the ADF image; note that inner empty space of the tube is detectable.

**Fig.1**

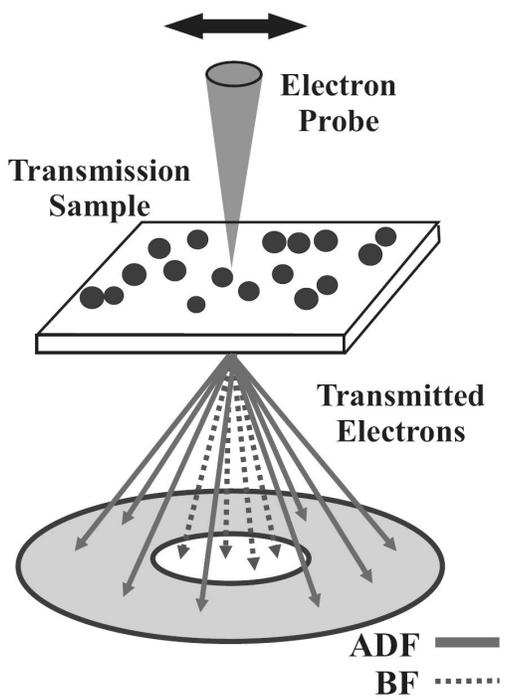

**Fig.2**

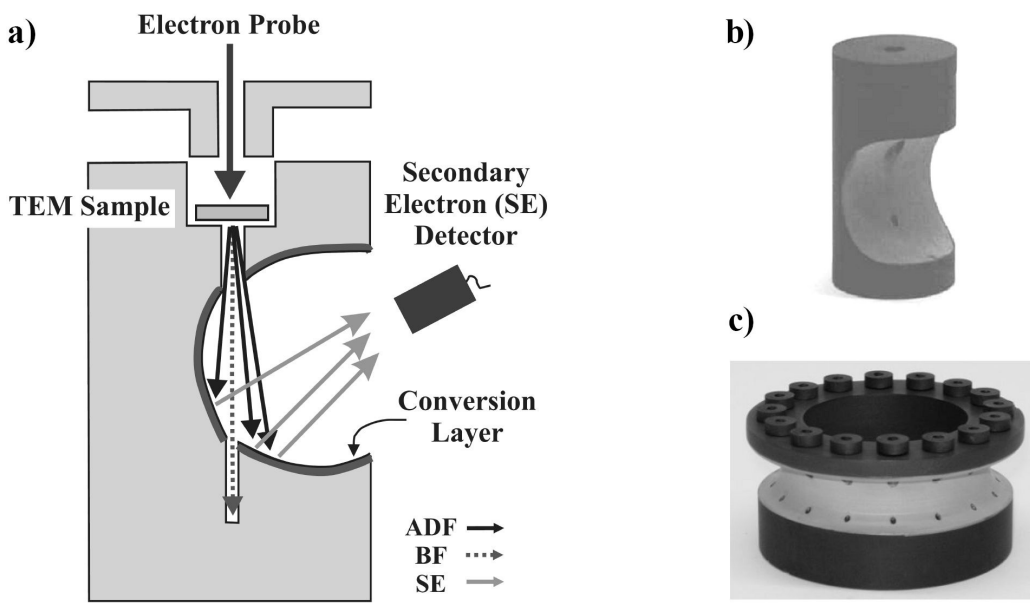

**Fig.3**

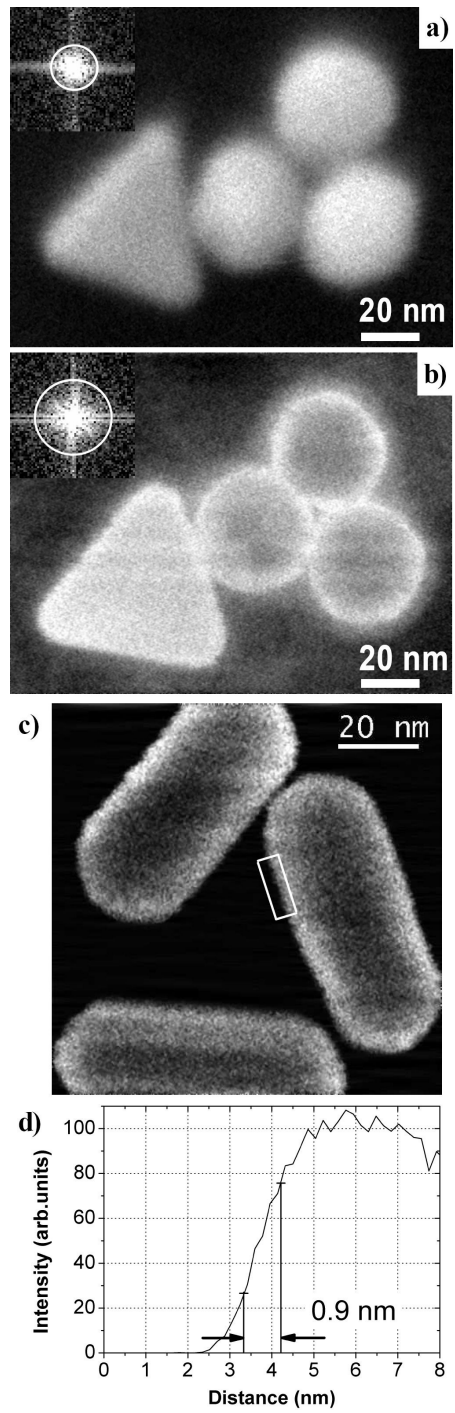

**Fig. 4**

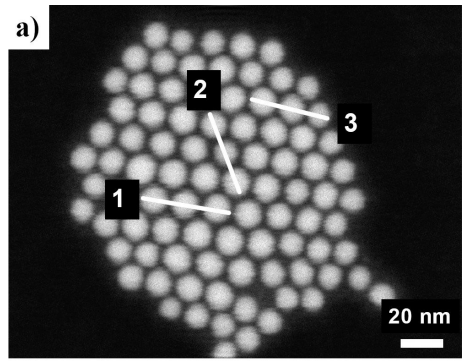

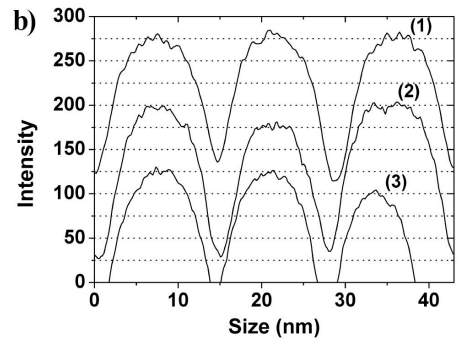

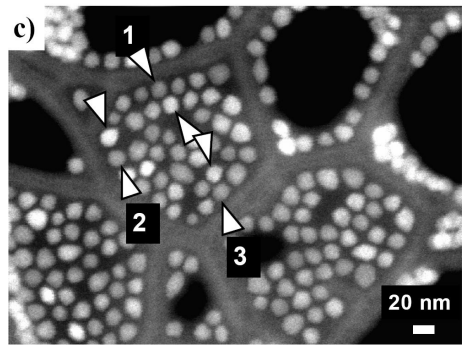

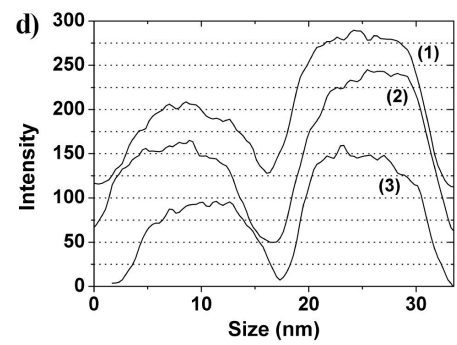

**Fig.5**

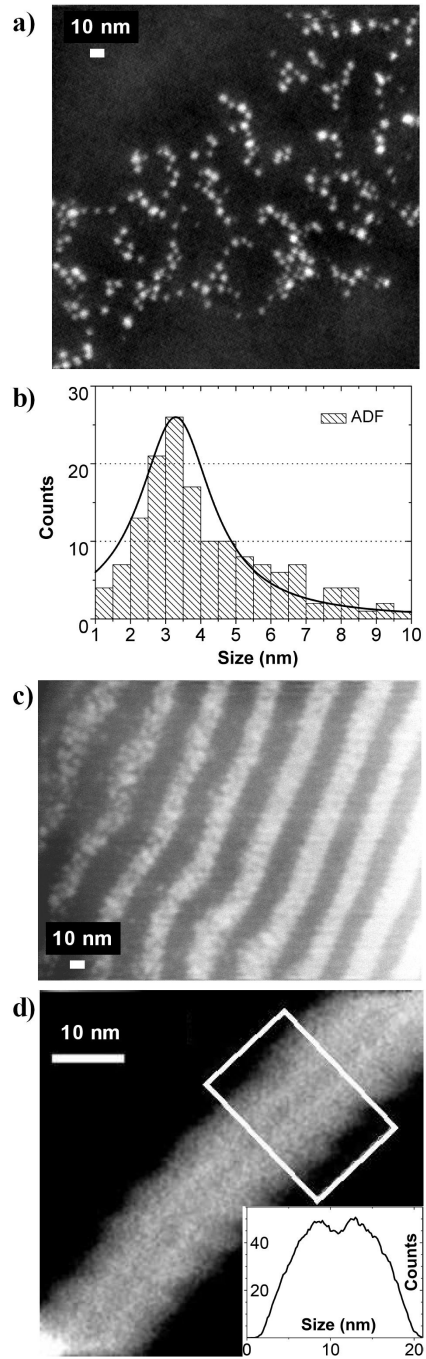